\documentclass[showpacs,twocolumn,pre,floatfix]{revtex4}
\usepackage{psfrag,epsfig,amsfonts,amssymb,amsmath}
\usepackage{dcolumn}

\newcommand{\bx}{{\bf x}}
\newcommand{\bu}{{\bf u}}
\newcommand{\bJ}{{\bf J}}
\newcommand{\bE}{{\bf E}}
\newcommand{\kB}{k_{\mathrm{B}}}
\newcommand{\dmin}{d}
\newcommand{\dmax}{D}
\begin{document}

%\title{Hydrostatic energy barriers suppress the 
%translocation of particles through nanopores}
%\title{Reluctance of an electrically neutral object to enter a charged nano-pore}
%\title{Reluctance of an neutral object to enter a charged nano-pore}
%\title{On the Reluctance of a neutral nano-particle to enter a charged pore}
%\title{Opposite translocation of short and long polymer fragments through a nanopore}
\title{Opposite translocation of long and short oligomers through a nanopore}

\author{Sebastian Getfert}
\author{Thomas T\"ows}
\author{Peter Reimann}
\affiliation{Universit\"at Bielefeld, Fakult\"at f\"ur Physik, 33615 Bielefeld, Germany}

\begin{abstract}
We consider elongated cylindrical particles,
modeling e.g. DNA fragments or nano-rods, 
while translocating under the action of an externally
applied voltage through a solid state nanopore.
Particular emphasis is put on the concomitant
potential energy landscape,
encountered by the particle on its passage through the pore
due to the complex interplay of various electrohydrodynamic effects
%electrophoresis, 
%electroosmosis, polarization effects,
%(self energy and charge polarization), 
%and osmotic pressure
%and counter ion induced NEIN osmotic pressure
%(by counter ions)
%and screening 
beyond the realm of small Debye lengths.
We find that the net potential energy difference 
across the membrane may be of opposite sign for short 
and long particles of equal diameters and charge densities (e.g. oligomers).
%,and can even be controlled by an additional membrane gate electrode. 
Thermal noise thus leads to biased diffusion through the 
pore into opposite directions.
By means of an additional membrane gate electrode
it is even possible to control the specific particle length
at which this transport inversion occurs.
%The specific particle length at which this transport inversion occurs
%can be controlled by means of a membrane gate electrode.
%they exhibit a biased diffsion through 
%the pore in opposite directions
%and hence they preferentially travel in 
%opposite directions through the pore in the 
%presence of thermal noise.
\end{abstract}

 % Theory, modeling, and computer simulation
% 83.50.Ha 	Flow in channels (see also 
          % 47.60.Dx Flows in ducts and channels in fluid dynamics)
%47.61.Fg Flows in micro-electromechanical systems (MEMS) 
          % and nano-electromechanical systems (NEMS)
%47.61.-k 	Micro- and nano- scale flow phenomena
%87.14.G- 	Nucleic acids
%87.14.gk 	DNA
%87.15.H- 	Dynamics of biomolecules
%87.50.ch 	Electrophoresis/dielectrophoresis and other 
           % mechanical effects (see also 87.15.Tt Electrophoresis)
\maketitle
%permeation
%electric double layer

\section{Introduction}
\label{s1}
The translocation of polyelectrolytes 
and other biomolecules through membrane pores 
and channels 
plays a prominent role in a wide variety of biological 
contexts, and has recently attracted much attention
as a new paradigm for single molecule 
%detection,
analysis and manipulation
%%e.g.opening promising perspectives regarding
%%future develpments towards 
like DNA sequencing and other medical diagnostic 
applications 
\cite{reviews}.
Generally speaking, and disregarding the 
often quite different underlying physics, 
%a key feature of a pore
one of the most remarkable features of 
%membranes 
a pore
%A key aspect of both biological (protein) and 
%artificial (solid state) pores 
%is the possibility of
is its selectivity regarding the permeability 
%selective permeability 
by different particle species 
\cite{select1}.
In its most pronounced form,
%An even more pronounced 
%%form of 
%selectivity,
namely translocation of different particle species
into opposite directions,
it has been explored 
in much detail 
e.g. in the context of particle sorting by
structured microfluidic channels
\cite{select2}.
%One of the most important features of 
%such membranes pores 
%a pore
%A key aspect of both biological (protein) and 
%artificial (solid state) pores 
%is its selectivity, either with respect 
%to its permeability by different
%particle species \cite{sel1},
%or even by giving rise to translocation
%into opposite directions for different 
%particle species 
%\cite{rou04,pnas??}.
%The latter, most extreme form of selectivity
%has been explored in much detail e.g. in the context
%of ion channels \cite{rou04??} 
%and for particle sorting on the micrometer scale 
%by structured microfluidic devices
%\cite{pnas,austin,eichhornetal,mueller and matthias...}.

The main objective of our present work
is to extend those ideas for the purpose 
of separating DNA fragments, nano-rods
etc. of different lengths by using a solid state nanopore.
We theoretically predict the possibility of 
opposite translocation directions in response to an 
externally applied, static voltage difference across 
the membrane.
Moreover, the specific particle length at which the transport 
inversion occurs can be readily modified by means of an
additional membrane gate electrode.
%\cite{he11,nam09,ai10,gra06,lua10} 
%into the membrane.
A crucial point is that all particles are assumed 
to exhibit the same diameter and the same charge 
density (per lengths unit or per surface area unit).
It is only the particle length which may differ and
which then may result in opposite translocation 
directions through the pore. In other words, 
given a particle of suitable length,
after breaking that particle into two equal
pieces, those two pieces will move
through the pore in the opposite direction 
than the original, long particle.

%One of the most important features of 
%such membranes pores 
%a pore
%A key aspect of both biological (protein) and 
%artificial (solid state) pores 
%is its selectivity, either with respect 
%to its permeability by different
%particle species \cite{sel1},
%or even by giving rise to translocation
%into opposite directions for different 
%particle species 
%\cite{rou04,pnas??}.
%The latter, most extreme form of selectivity
%has been explored in much detail e.g. in the context
%of ion channels \cite{rou04??} 
%and for particle sorting on the micrometer scale 
%by structured microfluidic devices
%\cite{pnas,austin,eichhornetal,mueller and matthias...}.
%The main emphasis of our present work is on
%the separation of DNA fragments, nano-rods
%etc. of different lengths by using a solid state nanopore.
%We theoretically predict the possibility of 
%opposite translocation directions in response to an 
%externally applied, static voltage difference across 
%the membrane.
%Moreover, the specific particle length at which the transport 
%inversion occurs can be readily modified by means of an
%additional membrane gate electrode.
%\cite{he11,nam09,ai10,gra06,lua10} 
%into the membrane.

\begin{figure}
\begin{center}
	\epsfxsize=0.85\columnwidth
	\epsfbox{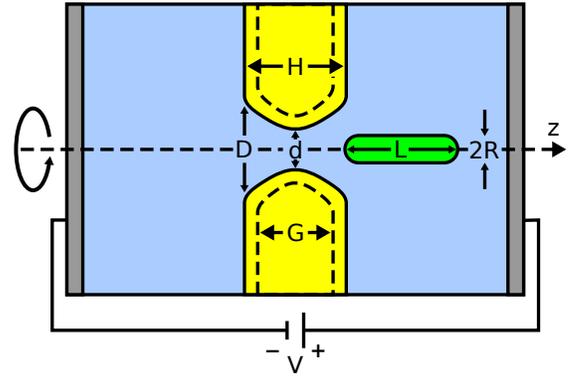} 
\end{center}
\caption{\label{fig1}
(Color online) Schematic illustration of the system.
A membrane of thickness $H$ (yellow) separates 
two compartments with electrolyte solution (blue)
and contains an hourglass-shaped nanopore 
with minimal and maximal diameters $\dmin$ and $\dmax$
\cite{f1}.
The dashed lines indicate the possibility of an 
additional gate electrode of thickness $G$, as 
considered later in the paper.
A voltage $V$ is applied to the electrodes (gray)
%(for more details see main text).
and acts on a prolate particle with radius $R$ 
and length $L$ (green).
%translocates through the pore. 
%along the z-axis.
% (dashed line). 
The system is symmetric about the $z$-axis 
and $z$ denotes the distance between pore and 
particle center.
%with pore center at $z=0$.
%The $z$-axis indicates rotation symmetry and
%the pore center is at $z=0$.
%The complete setup is rotationally symmetric about the 
%$z$-axis, whose origin ($z=0$) is at the pore center.
%For numerical convenience, the ``corners'' of
%pore and gate electrode are
%slightly rounded.
}
\end{figure}

%\section{Model}
%\label{s2}
A typical set-up we have in mind is 
sketched in Fig. \ref{fig1}. 
Such systems are governed by a complex 
and often quite non-intuitive
%and competition 
interplay of various electrohydrodynamic effects 
%\cite{gho07,he11,ai10,zha12,lua11,lua08,rab05,gro10,kes11,par69,elhyd,par73,bon06}
\cite{gro10,zha07,lua08,rab05},
which are mainly rooted in
%The most important ones are usually
%due to 
the electric double layers
%, which emerge 
at the membrane and 
particle surfaces 
due to 
%and which have their origin in surface charges 
%of the perforated membrane and the particle with 
certain, approximately constant 
surface charge densities $\sigma_m$ and 
$\sigma_p$, respectively 
\cite{sme06,gho07,dor09,he11}:
These charged surfaces attract counterions
(and repel coions) 
of the ambient electrolyte, resulting in a Debye screening layer 
of width $\lambda_D$ \cite{bru08}.
%To begin with, an electric double layer
%emerges at the interface between 
%electrolyte solution and perforated 
%membrane, consiting of an (approximately) 
%constant surface charge density $\sigma_m$
%and a counter-ion layer of
%width $\lambda_D$ (Debye length).
%Likewise, the particle exhibits a surface charge
%density $\sigma_p$, screened by a Debye layer of the
%same width $\lambda_D$ \cite{bru08}.
An externally applied voltage
%The external voltage in Fig. \ref{fig1}
generates forces on all those
fixed and mobile charges
%charged surfaces and counter-ion layers
and thus leads to 
an electroosmotic fluid 
flow superimposed by an electrophoretic 
motion of a ``free'' particle, or
to equivalent hydrodynamic
and electrostatic forces on an 
immobilized particle \cite{lua08,dor09,kes11}.
Since the membrane is insulating,
%the pore is much smaller than the fluid chambers, 
nearly the entire voltage drop and the 
hence induced forces actually occur 
in the nanopore and its immediate 
neighborhood \cite{zha07,dor09,gro10}.

%The resulting net force $F(z)$
%on a particle at an arbitrary but fixed 
%position along the $z$-axis in Fig. 1, 
%or, equivalently, the potential $U(z)$ 
%with $U'(z)=-F(z)$, is the quantity of 
%foremost interest in our present work.
%
%ev nicht
%In the following, we will first 
%provide an intuitive picture of the 
%main physical mechanisms contributing to $U(z)$.
%Then, we will turn to a quantitative 
%description in terms of the Poisson-Nernst-Planck 
%(PNP) and Stokes equations.
%Within this formalism, a more rigorous 
%justification of our intuitive reasoning
%is straightforward but quite tedious, and 
%hence is postponed to a later, more detailed 
%publication.
%Instead, we will confine ourselves to
%representative numerical solutions in order 
%to quantitatively corroborate our general 
%qualitative predictions.

In the next section we will provide an intuitive picture 
of the main physical mechanisms governing those forces.
Subsequently, we will turn to a quantitative 
illustration based on the Poisson, Nernst-Planck,
and Stokes equations (Sects. \ref{s3} and \ref{s4}).
Generalizations of the model,
in particular the effects of an additional membrane gate
electrode (dashed line in Fig. \ref{fig1}),
are covered by Sect. \ref{s5}.
Our symmary and conclusions are provided by Sect. \ref{s6}.

\section{Qualitative Considerations}
\label{s2}
We first focus on thin Debye layers \cite{fx},
i.e. $\lambda_D$ is much 
smaller than the distance between perforated 
membrane and 
immobilized particle, and also much smaller 
than any curvature-radii of their charged 
surfaces (cf. Fig. \ref{fig1}).
Disregarding for a moment those thin Debye 
layers, the remnant ``outer region'' of 
the electrolyte solution
%Denoting these thin Debye layers as the
%boundary region of the electrolyte 
%solution, the remnant ``outer region'' 
exhibits the following
``similitude'' property \cite{similitude}:
Whatever the electric field in the system 
from Fig. \ref{fig1} will look like,
the concomitant electroosmotic velocity field will be proportional
to this electric field throughout the outer region, 
%The electroosmotic velocity field is 
%proportional to the electric field 
%throughout the outer region, 
provided that membrane and particle exhibit identical zeta-potentials.
In our present case of thin Debye layers, 
the latter condition is tantamount to
%or -- since $\lambda_D$ is small --
identical surface charge densities, i.e. 
$\sigma_m=\sigma_p$ \cite{bru08}.
%In conjunction with
%Complementing this similitude property by 
%explicit solutions within the thin Debye 
%layers, this ``similitude'' 
This similitude property of the outer 
region can be complemented by means of matched 
asymptotic expansion techniques
with explicit solutions for the electric and
fluid flow fields within the thin Debye 
layers \cite{similitude}.
The result is an exact cancellation of
%Whence one readily can infer that 
the hydrodynamic and electrostatic forces
on any charged surface element.
%exactly cancel.
We thus arrive at the quite remarkable
%implies an exact cancellation of the
%hydrodynamic and electrostatic forces \cite{similitude}, 
%i.e.
conclusion that no net force 
(nor torque)
is acting on the particle in Fig. \ref{fig1}
for equal surface charge densities ($\sigma_m=\sigma_p$)
and asymptotically small Debye lengths $\lambda_D$,
independently of 
%any further details of the system!
the external trans-membrane voltage
and any further details (shape, layout, material)
of membrane, pore, and particle!

What happens beyond thin Debye layers?
First of all, constant zeta-potentials are then
no longer tantamount to constant surface 
charge densities.
In particular, the zeta-potentials 
now depend, amongst others, on 
%the applied voltage, 
the surface curvatures and the distance 
between particle and membrane 
\cite{bru08,yar12}.
Henceforth, we thus we adopt the latter, physically 
more natural assumption of (approximately) 
constant surface charge densities 
$\sigma_m=\sigma_p$ \cite{beh01,he11}.

In a first step, we consider a spherical
particle of radius $R$ in the above special 
case $\sigma_m=\sigma_p$ and $R\gg\lambda_D$.
How will the concomitant cancellation of
hydrodynamic and electrostatic forces 
be modified as $R$ decreases?
In the opposite limit $R\ll\lambda_D$ there
%We begin with a spherical particle
%in the above special case
%$\sigma_m=\sigma_p$ and thin 
%Debye layers.
%How do the forces change 
%as the particle radius $R$ decreases?
%In the limit $R\ll\lambda_D$ there
are negligibly few ions within
the hydrodynamically relevant 
neighborhood of the particle  \cite{bru08},
and thus the electrostatic force equals
charge $4\pi\sigma_p R^2$
times electric field,
%$\bf{E}$, 
while the hydrodynamic force equals Stokes friction
$6\pi\eta R$ ($\eta$ being the fluid 
viscosity) times fluid velocity
%$\bf{u}$ 
(caused by electroosmosis).
%In particular, along the $z$-axis in Fig. \ref{fig1}
%this amounts to a net force
%\begin{equation}
%F(z)=4\pi R^2\sigma_p\, E_z(z)+6\pi\eta R\, u_z(z) \ .
%\label{1}
%\end{equation}
%Returning to the general case, it is obvious
%that the hydrodynamic force ultimately
%``wins'' for sufficiently small $R$.
Due to the $R^2$ decay, the former becomes 
negligible as $R\to 0$.
It is furthermore quite plausible that the 
same trend
%, namely a predominance of electroosmosis, 
sets in as soon as $R$ becomes comparable
or smaller than $\lambda_D$,
i.e. the particle will be dragged into the 
direction of the electroosmotic flow.
In the most common case of negative surface 
charges \cite{beh01,he11},
this direction is from right to left
for the voltage sign convention of
Fig. \ref{fig1}.

Analogous conclusions readily carry over
to arbitrary particle shapes upon observing 
that when scaling down its size by a factor
$\alpha$, the surface and hence the charge
decrease as $\alpha^2$, whereas the Stokes 
friction decreases as $\alpha$.
However, while for spheres the predominance 
of electroosmosis sets in at $R\approx\lambda_D$,
the corresponding threshold will now depend on
the detailed particle shape.
E.g. for a cylindrical rod, whose radius $R$ is 
much smaller than its length $L$, 
the threshold is at 
$L\approx\lambda_D$ \cite{she82}.
For Debye lengths comparable 
to $R$ we thus expect that spheres and short 
rods are dragged by electroosmosis from right
to left in Fig. \ref{fig1}, 
while the net force on long rods is still negligible.
%while for long rods this effect is still negligible.

So far, we tacitly ignored
any spatial variations of the 
electric and fluid flow fields over the 
particle-length $L$, as well as any
back-coupling of the particle
to those fields.
%In the following, we argue that
%these finite particle-size effects in fact 
%may further reinforce the opposite translocation 
%directions of small and large particles 
%quite significantly.
%
%Practically, a growth of the Debye lengths
%can then the be readily realized e.g. by 
%reducing the ion concentrations \cite{bru08}.
%Typically, the first correction to the
%thin Debye layer asymptotics will be due 
%to the finite curvature of the particle 
%surface.
%For simplicity, we thus focus for a moment
%on a  spherical particle of radius $R$ 
%
To better understand those finite particle size effects,
we now turn to another simple limiting case, namely 
$\lambda_D\to\infty$.
%when the ion concentration tends to zero and thus 
%$\lambda_D\to\infty$ \cite{bru08}.
%when the Debye length tends to infinity.
%ev. nicht
%Put differently, we may say that non-negligible
%surface curvatures tend to diminish the 
%electrophoretics of a particle.
%Analogous tendencies may thus be expected 
%when the perforated membrane exhibit 
%non-negligible surface curvatures 
%(compared to $\lambda_D$):
%They should tend to diminish the 
%electroosmotics if they are positive 
%(convex) and vice versa if they
%are negative (concave).
%However, usually the particle curvatures
%are considerably larger anyway,
%and hence those modifications
%of the electroosmosis are minor effects.
%
%ev:
%For typical pore geometries, the positive
%curvatures prevail, hence giving rise
%to a competing tendency, especially
%for non-small particles, namely 
%to promote a net force on the resting 
%particle opposite to the electroosmotic 
%flow.
%
%Yet another quite obvious case arises
%for very large Debye lengths. 
%Once again, the system then behaves 
As expected \cite{fx}, the system then behaves 
essentially as in the absence of any ions: 
%\cite{bru08}:
Electroosmosis is negligible,
and upon entering the pore, the particle 
is strongly repelled by the hardly 
screened like-charges of the membrane, 
and symmetrically when exiting the pore.
From the energetic viewpoint,
the corresponding two energy 
barriers thus nearly cancel each other
so that there still remains an 
net potential energy gain when a 
negatively charged particles moves
from the left to the right fluid
compartment in the presence of
a voltage with the sign 
convention from Fig. \ref{fig1}.
Upon gradually reducing the Debye length,
the potential barriers as well 
as the net energy gain for left-to-right 
translocations will clearly decrease, 
but it is also quite plausible that 
we may still encounter remnants
of both effects down to quite small 
Debye lengths:
The barriers may then also be
viewed as due to the ``bumping'' 
of the counterion cloud (Debye layer)
around the particle into that
around the membrane, and the
overall predominance of the electrostatic
(from left to right) 
over the hydrodynamic forces 
(from right to left)
may be viewed as due to the fact
that a merging of the counterion clouds 
%expels some of thecounterions into regions with
reduces the number of counterions 
%brings along a reduction of counterions
in regions with high external fields
%brings along an overall reduction of the 
%total number of counterions 
and thus the 
%%electroosmotic 
fluid flow which they induce.
Finally, it is also quite clear 
that for any given Debye length,
both effects (energy barriers and 
predominance of electrostatic forces)
will become weaker and weaker
as the particle size decreases.

Altogether, we thus predict that for 
$\sigma_m\approx \sigma_p$,
$\lambda_D \approx R$, and pore diameters of a few $R$,
there will be a competition between the
two above mentioned effects, the first
prevailing for short and the second for long rods.
Hence, the potential energy difference across the same pore
will be of opposite sign for long and short rods.

\section{Numerical Treatment}
\label{s3}
To quantitatively
%Below we will quantitatively 
%verify this heuristic prediction for 
verify our above predictions, we 
%focus on the following representative example:
%As in Fig. \ref{fig1}, we 
consider the system from Fig. \ref{fig1} with
a cylindrical fluid chamber, a planar, solid state 
membrane of thickness $H=20\,$nm,
surface charge density 
$\sigma_m=-50\,$mC/m$^2$ \cite{beh01,he11},
and an hourglass-shaped pore 
with $\dmin=10\,$nm and $\dmax=20\,$nm \cite{f1}.
The electrolyte's temperature and viscosity are
$T=300\,$K and  $\eta = 10^{-3}\,$Pa$\,$s,
and we assume two ionic species with 
opposite charges
$q_1=-q_2=e=1.6\ ...\cdot 10^{-19}\,$C,
equal thermal 
diffusion coefficients 
$D_1=D_2=2\cdot10^{-9}$ m$^2$/s,
%$D_1=D_2=:D=2\cdot10^{-9}$ m$^2$/s 
%$\diff=2\cdot10^{-9}$ m$^2$/s
%and bulk concentration
and equal bulk concentrations 
%($c_{1,bulk}=c_{2,bulk}=:c_{bulk}$),
$c_0=100\,$mM \cite{pro03}.
%%where $1$ mM denotes $10^{-3}$ mol per cubic meter.
%%The temperature is chosen as $T=300$ K 
%%the corresponding viscosity of water as
%%$\eta = 10^{-3}$ Pa\;s.
%An external voltage $V$ can be applied as 
%indicated in Fig. \ref{fig1}.
%%via two electrodes 
%%which are located 
%%at the side walls of the fluid chamber.
%%so that
%%As expected,
%%Because the membrane is insulating and because the radius of 
%%the pore is much smaller than that of the fluid chamber, 
%%nearly the entire voltage drop occurs in the 
%%nanopore and its immediate neighborhood.
Accordingly, the Debye length is 
%implying 
$\lambda_D\approx 1\,$nm \cite{bru08}.
The cylindrical particle of radius $R$ and
length $L\ge 2R$ (cf. Fig. \ref{fig1}) 
exhibits spherical caps
and becomes a sphere for $L=2R$.
%In our model the particle is a cylinder with half 
%spheres merged to the ends, 
%and hence the particle becomes a spheres if $L=2R$.
Besides actual nano-rods, we mainly have
in mind fragments of a polyelectrolyte
(e.g. DNA) which are sufficiently short
(oligomers) that our rigid particle 
remains a reasonable approximation 
\cite{he11,ai10,kes11}.
%Similarly, we assume that the ``true'' particle
%charges can be approximated by a
In the same vein, we approximate the
real particle charges by a 
constant surface charge density 
$\sigma_p$ \cite{gho07,dor09,che10}.
Henceforth, we specifically focus on the
radius $R=1.1\,$nm of double 
stranded DNA and lengths $L$ 
between $2R$ and $50\,$nm.
Given a DNA basepair-distance of about 
$0.3\,$nm, a nominal charge of $-2\,e$ per 
basepair, and a typical screening factor 
of about $60\%$ due to 
%Manning's counterion condensation
counterion adhesion (transient binding)
%\cite{gho07,he11,zha12,lua08,rab05,par73,kow12},
\cite{gho07,he11,lua08,rab05,par73,kow12},
the equivalent surface charge density 
%\cite{gho07,dor09,che10}
$\sigma_p\approx -50$ mC/m$^2$ is 
quite similar to $\sigma_m$.
%almost the same as that of the membrane. 
%(and pore).
Since equal particle and membrane surface
charges densities also represents 
%Since $\sigma_m=\sigma_p$ is also 
%conceptually most interesting,
%we first focus on this case. 
the conceptually most interesting situation,
%(see above), 
we first focus on this case. 
%($\sigma_m=\sigma_p=-50\,$mC/m$^2$) in what follows.
%
%As it turns out, the most interesting case 
%arises when this surface charge is equal
%to the surface charge $\sigma$ of the 
%membrane, and hence 
%we always focus on this case henceforth.
%
%For the sake of completeness, 

\begin{figure}
\begin{center}
\epsfxsize=1.0\columnwidth
\epsfbox{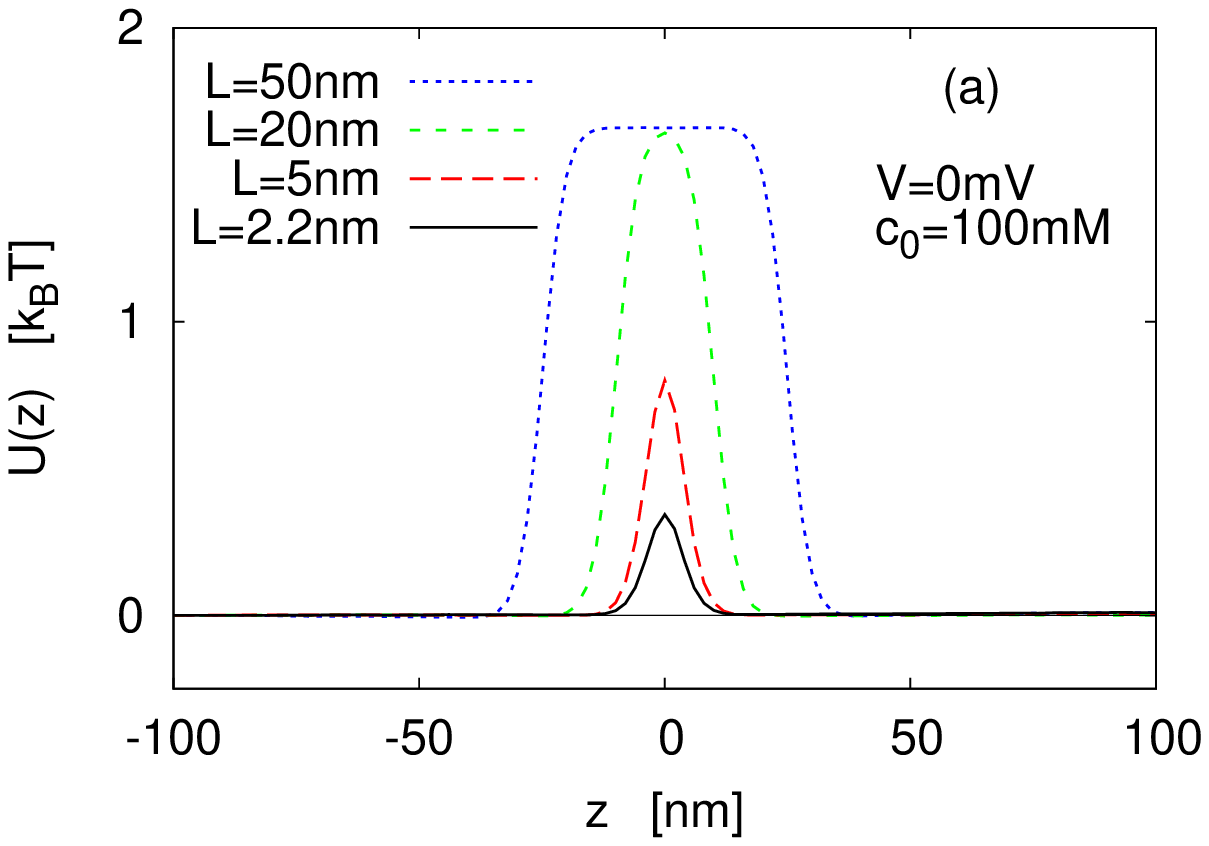} 
\epsfxsize=1.0\columnwidth
\epsfbox{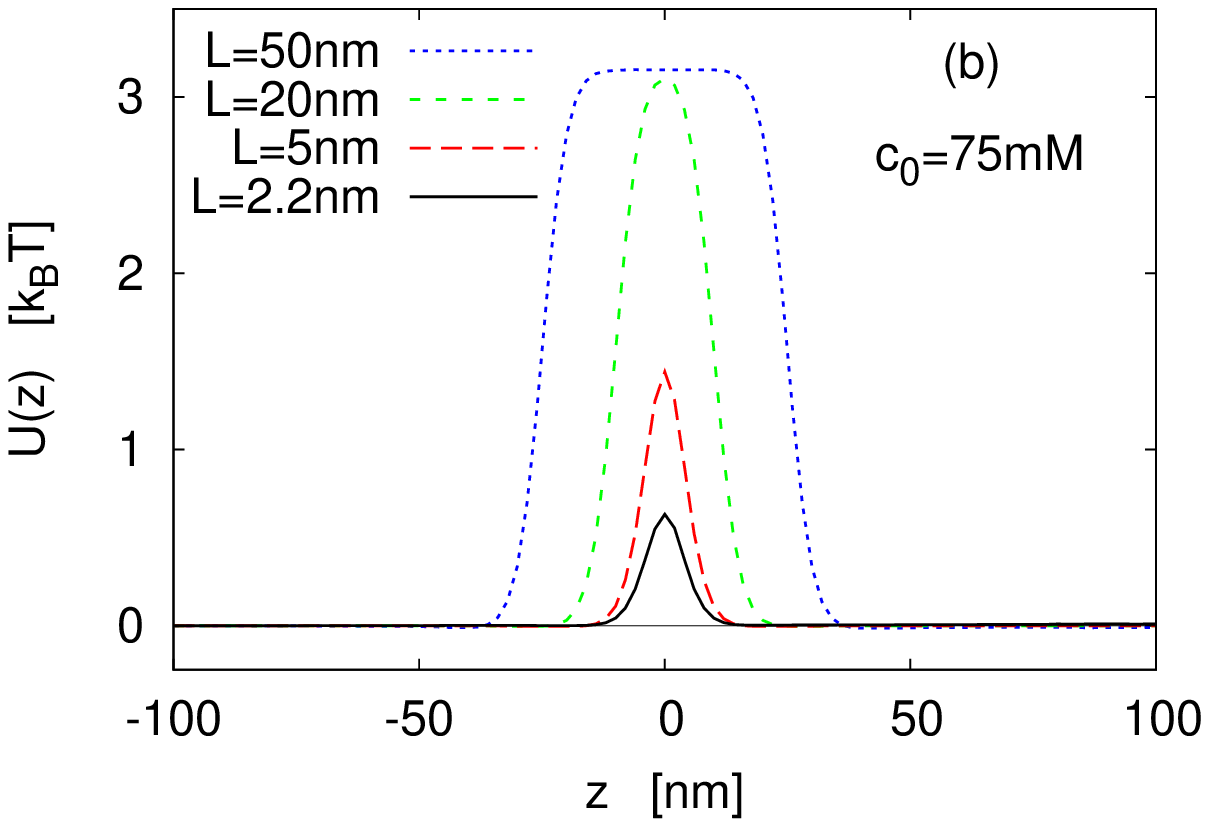} 
\end{center}
\caption{\label{fig2}
(Color online)
The potential energy $U(z)$ in units of 
the thermal energy 
$\kB T$ 
of a particle at position $z$ (in units of nm; cf. Fig. \ref{fig1})
%without external voltage ($V=0$) and parameters 
for different particle lengths $L$,
$V=0\,$mV, 
$c_0=100\,$mM, and
$\sigma_m=\sigma_p=-50\,$mC/m$^2$.
(a): Bulk concentration $c_0=100\,$mM.
(b): Reduced bulk concentration $c_0=75\,$mM.
%$H=20\,$nm,
%$\dmin=10\,$nm,
%$\dmax=20\,$nm, 
%%\cite{f1},
%$R=1.1\,$nm,
%and four $L$-values.
%$L=2.2\,$nm (black),
%$L=5\,$nm (red),
%$L=20\,$nm (blue),
%$L=50\,$nm (green).
%Dashed: Two examples ($L=2.2\,$nm and $L=50\,$nm)
%with reduced bulk concentration $c_0=75\,$mM.
%For further details see main text.
}
\end{figure}

%Finally, we will also
%account for polarization effects within 
Polarization effects within the electrolyte solution 
are taken into account via its 
dielectric constant (permittivity) $\epsilon_s$
\cite{jac99},
and likewise for membrane ($\epsilon_m$)
and particle ($\epsilon_p$).
Under the name ``self-energy'' they 
play a key role e.g. in narrow 
protein channels 
\cite{par69,zha07}, while for 
our present, much larger pores they
are expected to be of minor importance \cite{kes11}.
%Following \cite{par69,zha07,bon06},
We usually set
$\epsilon_s = 80\epsilon_0$,
$\epsilon_m = \epsilon_p = 5\epsilon_0$, 
where $\epsilon_0$ is the vacuum permittivity,
and we
verified that variations of $\epsilon_m$ and 
$\epsilon_p$ between $\epsilon_0$ and $\epsilon_s$
indeed only lead to minor changes e.g. in 
Figs. \ref{fig2} -- \ref{fig3a} \cite{f4}.
%, while the choice of $\epsilon_p$ 
%matters even much less.

Next we turn to the Poisson, Nernst-Planck, 
and Stokes equations, which are well-established
in this context \cite{eis96,pro03,mas06},
and thus only briefly summarized here.
Throughout the electrolyte solution
%, a constant permittivity $\epsilon_s$ 
%is assumed and 
the electric potential $\psi$ 
satisfies Poisson's equation 
$\epsilon_s \Delta \psi(\bx) = - \rho(\bx)$,
where 
%$\epsilon_0$ is the vacuum permittivity and
$\rho(\bx) = F_c\,[c_1({\bf x})-c_2({\bf x})]$
is the charge density in terms of
Faraday's constant $F_c$ and the molar 
concentrations $c_{1,2}(\bx)$
of the two ionic species (see above).
Likewise, $\Delta \psi(\bx) = 0$
throughout the perforated membrane and 
the particle.
%Likewise, $-\epsilon_m \Delta \psi(\bx) = 0$
%within the membrane and  
%$-\epsilon_p \Delta \psi(\bx) = 0$
%within the particle.
%Typical values are \cite{elhyd,bon06}
%$\epsilon_s = 80\,\epsilon_0$, 
%$\epsilon_m = 5\,\epsilon_0$, and
%$\epsilon_p = 5\,\epsilon_0$ \cite{f2}, 
%where
%%for solid-state membranes
%$\epsilon_0$ is the vacuum permittivity.
%%Since $\epsilon_p$ turns out to play a 
%%very minor role, we focus on
%%%Therefore, and in analogy to the surface 
%%%charges, we focus on the specific choice
%%%$\epsilon_p=\epsilon_m$.
%%$\epsilon_p=\epsilon_m$.
Boundary conditions are $\psi=\pm V/2$ at the 
electrodes and ${\bf n}\nabla \psi= 0$ 
at the cylindrical chamber walls 
(cf. Fig. \ref{fig1}),
where ${\bf n}$ indicates the surface normal.
At the charged liquid-membrane interface we 
require that 
${\bf n} [\epsilon_s \bE_s -\epsilon_{m}\bE_m]
=\sigma_m$ 
%(see e.g. Eq. (4.40) in \cite{jac99}),
where $\bE_s$ and $\bE_m$ are the electric
fields $-\nabla\psi$ at the two sides
of the interface \cite{ai10,kes11,yar12,jac99},
and analogously at the particle-liquid 
interface.

According to Nernst-Planck,
the flux densities $\bJ_\nu$ of the two
ionic species ($\nu=1,2$) 
%with equal diffusion coefficients $D$ 
%and charges $q_1=-q_2$ (see above) 
%are expressed 
%according to the Nernst-Planck equation
%in terms of the fluid velocity $\bu$,
%the electric potential $\psi$, and the molar
%concentrations $c_\nu$ as
are given by
$c_\nu(\bx) \bu(\bx) - D_\nu\nabla c_\nu(\bx) 
- \mu_\nu c_\nu(\bx) \nabla \psi(\bx)$,
where $\bu$ is the fluid velocity,
$\mu_\nu:=q_\nu D_\nu/\kB T$ the ion mobility, 
and $\kB$ Boltzmann's constant,
and they satisfy 
%the steady state continuity equation
%Particle number conservation 
%%implies a continuity equation, which 
%%simplifies under steady state conditions to 
%under steady state conditions implies
$\nabla \bJ_\nu(\bx) = 0$.
%%and which is complemented by the 
%%following boundary conditions:
At the 
%left and right walls (electrodes), 
electrodes, the concentrations $c_\nu$ 
must assume their bulk value $c_0$.
%(see above).
On all other 
%solid-liquid interfaces 
boundaries we impose
%(membrane, particle, cylindrical chamber wall) we adopt
%reflecting boundary conditions 
%are imposed on $\bJ_\nu$.
${\bf n}\bJ_\nu=0$.
%since the ions cannot enter into those regions.

%Focusing on steady states, 
The velocity $\bu$ and pressure $p$ 
%of the electrolyte solution 
satisfy Stokes equation
%the quasistatic Stokes equations
%(i.e. neglecting the non-linear 
%terms and gravity in the Navier-Stokes equation)
$\eta \Delta \bu(\bx) = \nabla p(\bx) + \rho(\bx) \nabla \psi(\bx)$
%and the continuity equation for incompressible fluids 
and $\nabla \bu(\bx) = 0$ with
no-slip boundary conditions $\bu = {\bf 0}$ 
%are adopted 
on membrane, particle, and cylindrical
chamber walls.
%(see Fig. \ref{fig1}).
At the electrodes we require:
%The remaining boundary conditions 
%at the side walls are:
(i) The pressure $p$ assumes a preset 
``bulk value'', whose actual choice turns 
out to be irrelevant.
%$p_0$. Its actual choice
%turns out to be irrelevant, hence we set $p_0=0$.
%which can be set equal to zero 
%since only $\nabla p$ actually matters.
(ii) The hydrodynamic stress tensor 
$A$ with matrix elements 
$A_{ij}:=\eta(\partial u_i/\partial x_j+\partial u_j/\partial x_i) - p\delta_{ij}$
satisfies
$A {\bf n}= {\bf 0}$.
%$A (\bx)\,{\bf n}(\bx) = {\bf 0}$.
These boundary conditions 
%(i) and (ii) 
are well-known to be 
numerically very efficient and stable \cite{mas06}.

Finally, the force ${\bf F}$ 
on the 
%immobilized 
particle is obtained
as the integral $\int_S [A(\bx)+B(\bx)]\,
{\bf n}(\bx) dS$ over the particle surface $S$  \cite{mas06},
where 
%$A$ is the hydrodynamic stress tensor from above, and
$B$ is the Maxwell 
stress tensor with matrix elements 
$B_{ij}:=\epsilon_s(E_i\, E_j - \delta_{ij} |\bE|^2/2)$
and
$\bE  := -\nabla \psi$.
Focusing on particle positions along the $z$-axis
(cf. Fig. \ref{fig1}), we denote by $z$ 
the distance between pore- and particle-center,
%its center of mass 
and by $F(z)$ the $z$-component of 
${\bf F}$.
%at any given position $z$ of the particle center 
%(the other components vanish for symmetry reasons).
The potential energy $U(z)$ then follows
by integrating $-F(z)$ and setting $U(z)=0$ 
when the particle touches the left electrode
in Fig. \ref{fig1}.

Below, we present numerical
results for $U(z)$ obtained along these lines
with the COMSOL 4.3a finite element package of coupled
partial differential equation solvers \cite{f2}.
%In the following we present results
%obtained with the COMSOL Multiphysics 
%package implementation of
%partial differential equation solvers, 
%based on finite element methods.
We tested the numerics by verifying various
%The numerics was tested against various
analytically known special cases, 
including our above mentioned
%, quite remarkable 
prediction 
%\cite{similitude} 
that $U(z)\equiv 0$ for 
%$c_0\to\infty$ (implying 
$\lambda_D\to 0$ \cite{fx}.
% \cite{bru08}).
%independently of the voltage $V$, the surface charge
%density $\sigma$, and the particle- and pore-size.
Furthermore, we made sure that finite-size 
effects of the fluid chamber in Fig. \ref{fig1}
remain negligible 
%for all our results below
by choosing a diameter of $200\,$ nm
and a length of $400\,$nm.
%In particular, nearly the entire voltage 
%drop was observed to occurs, as expected, 
%in the nanopore and its immediate neighborhood.

\section{Results}
\label{s4}
Without external voltage,
Fig. \ref{fig2} shows results for the 
above specified parameter values.
As expected, upon entering the pore from 
either side, the particles encounter 
%quite notable 
potential barriers, which decrease for
smaller particles and increase with increasing 
Debye length $\lambda_D$ \cite{fx}.
%(tantamount to decreasing $c_0$ \cite{bru08}).
%Comparison with results after ``switching off'' 
%the surface charge of the membrane or of the 
%particle (not shown) suggests that those barriers 
%are mainly due to the only partially screened 
%repulsion between the particle 
%and membrane charges
%(the Debye length is about $0.7\,$ nm).
%Further contributions 
%are due to well-known self energy,
%osmotic pressure, and charge polarization 
%effects \cite{par69,elhyd,par73,bon06,gro10,kes11}.
%%The ``plateaux'' of $U(z)$ for the long
%particles in Fig. \ref{fig2} (green and blue curves)
%may be heuristically explained by the approximate
%translation invariance when a 
%long particle ``fully threaded'' through the pore.
%Accordingly, the plateau widens
%and its height remains constant
%with increasing particle length.
%Likewise, the ``plateaux'' for the short
%particles (black lines Fig. \ref{fig2})
%reflect an approximate translation
%invariance when they are ``fully inside'' the pore.
%Accordingly, the plateau rises in height
%and shrinks in width as the particle grows.
%The transition between the two cases
%occurs when particle and
%pore have similar lengths (red line Fig. \ref{fig2}).
The ``plateau'' of $U(z)$ for $L=50\,$nm
%may be heuristically explained by the 
is due to an approximate translation invariance 
%of a short particle ``fully inside the pore'' or 
when both rod-ends stick far out of the pore \cite{lub99}.
%``fully threaded through the pore''.
%Accordingly,
%In the first case the plateau rises in height
%and shrinks in width as the particle grows
%(black and red lines).
%In the second case, the plateau widens
%and its height remains constant (blue and green lines).
%The transition occurs when particle and
%pore have similar lengths (red line).
%With decreasing concentration $c_0$
%(increasing Debye length), the 
%potential barriers in Fig. \ref{fig2} 
%grow, as expected (dashed lines).
Accordingly, one finds that for even longer rods 
the width of the plateau grows, while 
its height and
the shape of the ``potential steps'' at its edges
hardly change.

\begin{figure}
\begin{center}
\epsfxsize=1.0\columnwidth
\epsfbox{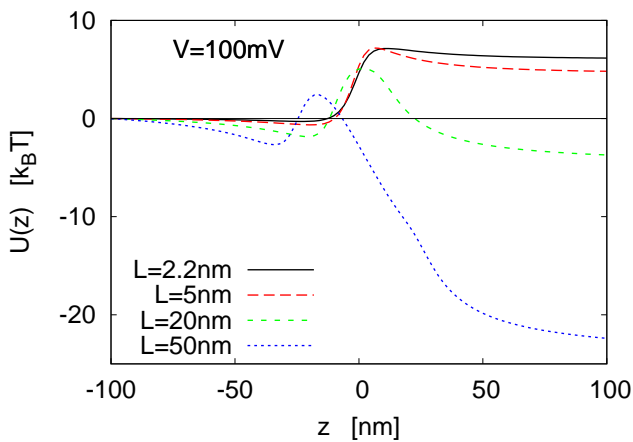} 
\end{center}
\caption{\label{fig3}
(Color online) Same as in Fig. 
\ref{fig2}(a) 
($c_0=100\,$mM, $\sigma_m=\sigma_p=-50\,$mC/m$^2$)
but now for an 
externally applied voltage of $V=100\,$mV.
}
\end{figure}

Typical results for finite voltage
are depicted in Fig. \ref{fig3}:
\cite{f3}:
%A typical example with finite external voltage
%is depicted in Fig. \ref{fig3} (solid lines):
Very roughly speaking, the barriers 
from Fig. \ref{fig2} get ``tilted'' 
by the external voltage.
However, this ``tilt'' is very different 
for different particle lengths, and neither
spatially homogeneous nor symmetric 
about $z=0$.
Only for $L=50\,$nm, a
region with constant slope 
develops around $z=0$, analogous to 
the ``plateau'' in Fig. \ref{fig2} \cite{lub99}.
Moreover, the potentials $U(z)$ 
approach their large-$z$ limits 
much slower than in Fig. \ref{fig2}
%in accordance with Ref. \cite{gro10}.
and in agreement with the $1/|z|$ asymptotics 
of $\psi$ predicted in \cite{gro10}.
Most importantly, the potential energy 
difference $U(z)-U(-z)$ for large $z$
is indeed of opposite sign for short and 
long particles, 
%in agreement with our prediction.
and also the signs themselves agree with our predictions
from Sect. \ref{s2}.

Note that $U(z)$ in Fig. \ref{fig3} is not monotonically
decreasing (increasing) for long (short) rods.
However, the intermediate potential barriers 
%in the pore-region 
of a few $\kB T$ are readily surmountable 
by thermal noise.
%Figs. \ref{fig2} and \ref{fig3} thus
%predict biased diffusion of long
%and short particles
%into opposite directions!
We remark that higher voltages $V$ lead to
larger potential energy differences \cite{f3}
and hence a better selectivity of the pore, 
but also the barriers become less easy 
to surmount.

Since the externally applied voltage drop $V$ 
mainly occurs within and close to the pore 
(see above and end of Sect. \ref{s1}),
the concomitant ``purely electrostatic'' transmembrane
potential energy difference follows as $qV$, where 
$q$ is the total particle charge, i.e. the 
net particle surface times its surface charge 
density. The actual potential
energy differences in Fig. \ref{fig3}, 
i.e. $U(z)-U(-z)$ for $z\geq 100\,$nm, 
is considerably smaller than $qV$ for the long 
particles ($L=50\,$nm and $L=20\,$nm)
and even of opposite sign for the short particles.
The main reason for those differences are clearly
the electroosmotic forces (cf. Sect. \ref{s2}),
which in fact would exert forces even when the 
particles were not charged at all (i.e. $qV=0$).

\begin{figure}
\begin{center}
\epsfxsize=1.0\columnwidth
\epsfbox{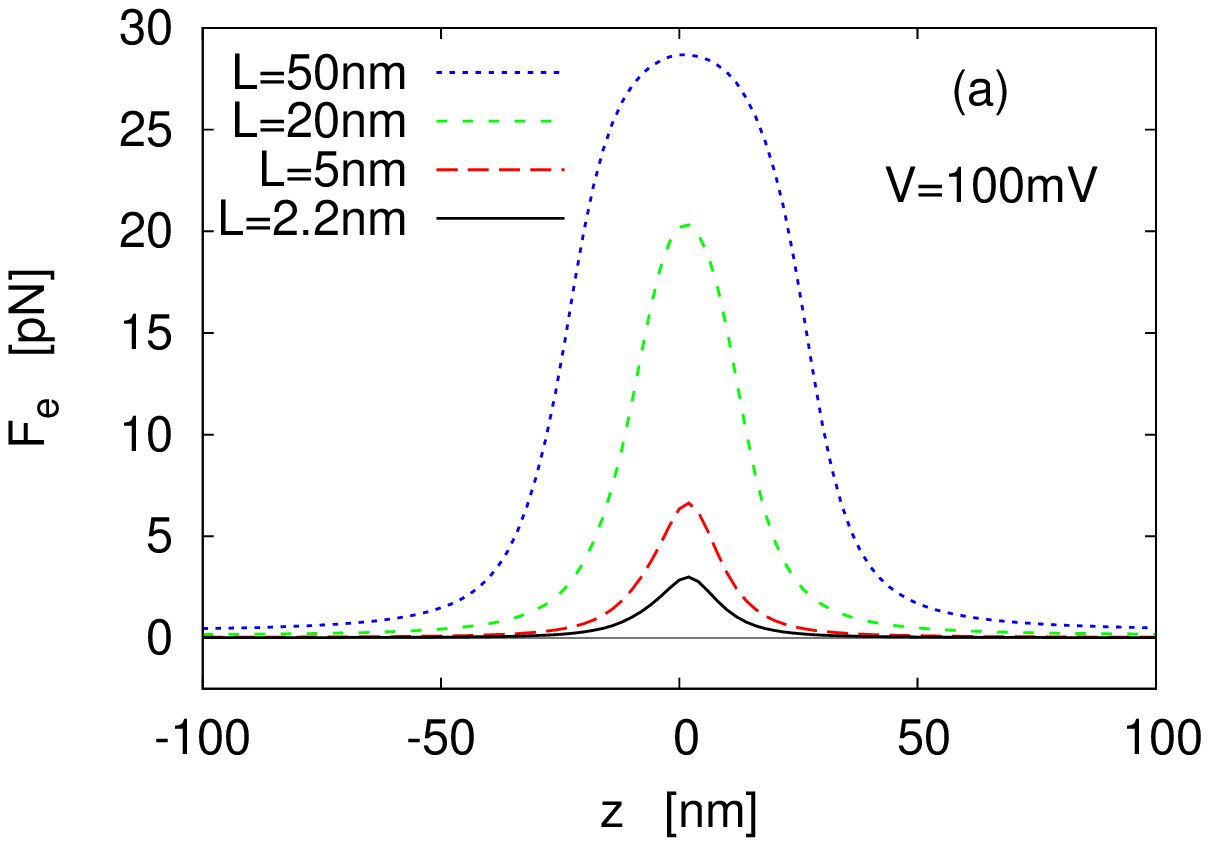} 
\epsfxsize=1.0\columnwidth
\epsfbox{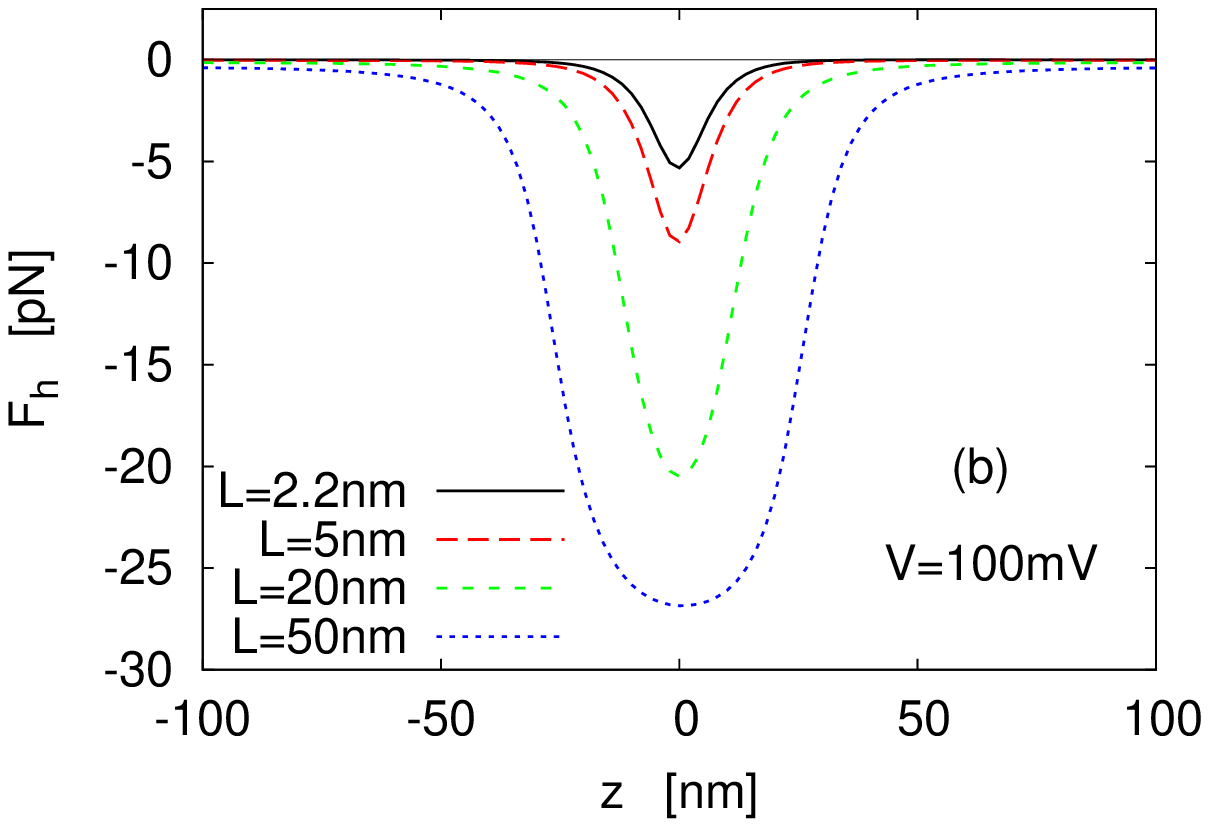} 
\end{center}
\caption{\label{fig3x}
(Color online) The two partial forces $F_e(z)$ and $F_h(z)$
in units of pN versus z in units of nm,
obtained (as detailed at the end of Sect. \ref{s3})
by integration of the Maxwell and hydrodynamic
stress tensors over the particle surface, respectively,
for the same system as in Fig. \ref{fig3}.
In other words, the potentials $U(z)$ from Fig. \ref{fig3}
are recovered upon integration of
$-F_e(z)-F_h(z)$.
}
\end{figure}

Finally, the electric and hydrodynamic force
contributions to the net potentials from 
Fig. \ref{fig3} are exemplified by Fig. \ref{fig3x}.
As detailed at the beginning of Sect. \ref{s3},
for asymptotically small Debye lengths, those two partial 
forces cancel exactly. Accordingly, going beyond small 
Debye lengths is essential to obtain non-vanishing
net forces (and potentials).
Yet, our Debye lengths are still so small that 
the two partial forces remain quite 
similar in modulus but of opposite sign in 
Fig. \ref{fig3x}.
In other words, they still almost 
cancel each other and the resulting
net force is much smaller than each 
partial force.
Note that while the forces in Fig.\ref{fig3x} 
are only slightly asymmetric about $z=0$, the
hence resulting asymmetry of the potentials 
in Fig. \ref{fig3} is more pronounced.

\section{Extensions}
\label{s5}
\begin{figure}
\begin{center}
\epsfxsize=1.0\columnwidth
\epsfbox{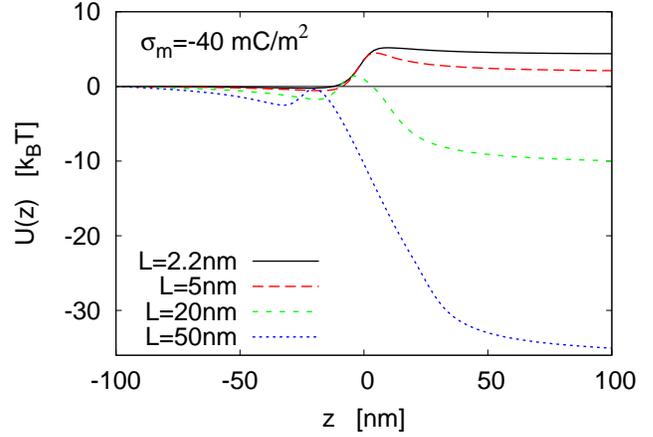} 
\end{center}
\caption{\label{fig3a}
(Color online) Same as in Fig. 
\ref{fig3} ($V=100\,$mV, $c_0=100\,$mM)
but now for an increased membrane
surface charge density of 
$\sigma_m=-40\,$mC/m$^2$.
}
\end{figure}

As a first generalization, 
we turn to unequally charged 
particle and membrane surfaces, say 
$\sigma_p=-50\,$mC/m$^2$ and
$\sigma_m=-40\,$mC/m$^2$.
%\cite{beh01,he11}.
%As an example, we consider
%a reduced particle charge of
%$\sigma_p=-40\,$mC/m$^2$,
%while the membrane is kept at
%the standard value $\sigma_m=-50\,$mC/m$^2$ 
%\cite{beh01,he11}.
%For thin Debye layers (large $c_0$), 
For small Debye lengths $\lambda_D$ \cite{fx},
the above mentioned similitude 
argument thus breaks down.
Rather, since $|\sigma_p|>|\sigma_m|$,
the hydrodynamic forces are now
overwhelmed by the electrostatic forces,
%electrophoresis (governed by $\sigma_p$) 
%overwhelms electroosmosis (governed by $\sigma_m$), 
i.e. all particles move preferentially
from left to right in Fig. \ref{fig1}.
Analogous modifications are expected for 
%our previous results for equal 
%surface charge densities and 
moderate Debye lengths.
%will now be superimposed by a systematic left-to-right preference.
This is confirmed by Fig. \ref{fig3a}, whose
main difference from Fig. \ref{fig3} is
%that all potentials $U(z)$ 
%now indeed exhibit 
an additional ``tilt to the right''.
Similarly, one finds that upon further 
increasing $\sigma_m$, this ``tilt to 
the right'' grows, and above about $-10\,$mC/m$^2$
all particles move preferentially from 
left to right.
Analogous effects are recovered for 
$\sigma_m<\sigma_p$ and upon variation
of $\sigma_p$ at fixed $\sigma_m$.

Finally, we will show that practically the 
same effects as by changing $\sigma_m$ can 
be generated by a membrane gate voltage.
Similarly as in Refs. \cite{he11,ai10,gate},
a gate electrode of thickness $G$
is integrated into the membrane
(dashed lines in Fig. \ref{fig1}),
and is thus still coated by a non-conducting 
layer of thickness $h:=(H-G)/2$.
The applied voltage $V_{gate}$ fixes the electric 
potential $\psi$ of the gate electrode relatively 
to the potentials $\psi=\pm V/2$ of the external 
electrodes.
%%Formally, within the gate electrode,
%Accordingly, Poissons equation is 
%%thus 
%replaced by 
%$\psi(\bx)=V_{gate}$, where $V_{gate}$ 
%is the externally applied gate voltage.
In the absence of the pore, one can 
analytically show that a gate voltage 
$V_{gate}$ is approximately equivalent to changing
the  ``bare'' membrane surface charge density 
$\sigma_m$ by $\Delta\sigma_m\approx\epsilon_m V_{gate}/h$.
%with $\chi=\epsilon_m V_{gate}/h$.
Numerically, we found that this 
equivalence remains valid in 
remarkably good approximation 
even in the presence of the pore.
For instance, for the system resulting 
in Fig. \ref{fig3},
an additional membrane gate electrode 
with $h=2\, $nm and $V_{gate}=500\,$mV
amounts to 
%$\chi\approx 20\,$mC/m$^2$V and
$\Delta\sigma_m\approx10\,$mC/m$^2$,
and indeed reproduces the dotted lines
quite well.

\begin{figure}
\begin{center}
%    \hspace*{-0.5cm}	
%	\epsfxsize=0.52\columnwidth
%	\epsfbox{figure4a.eps}
%	\hspace*{-0.5cm}
%	\epsfxsize=0.52\columnwidth
%	\epsfbox{figure4b.eps}%\\[0.3cm]
\epsfxsize=1.0\columnwidth
	\epsfbox{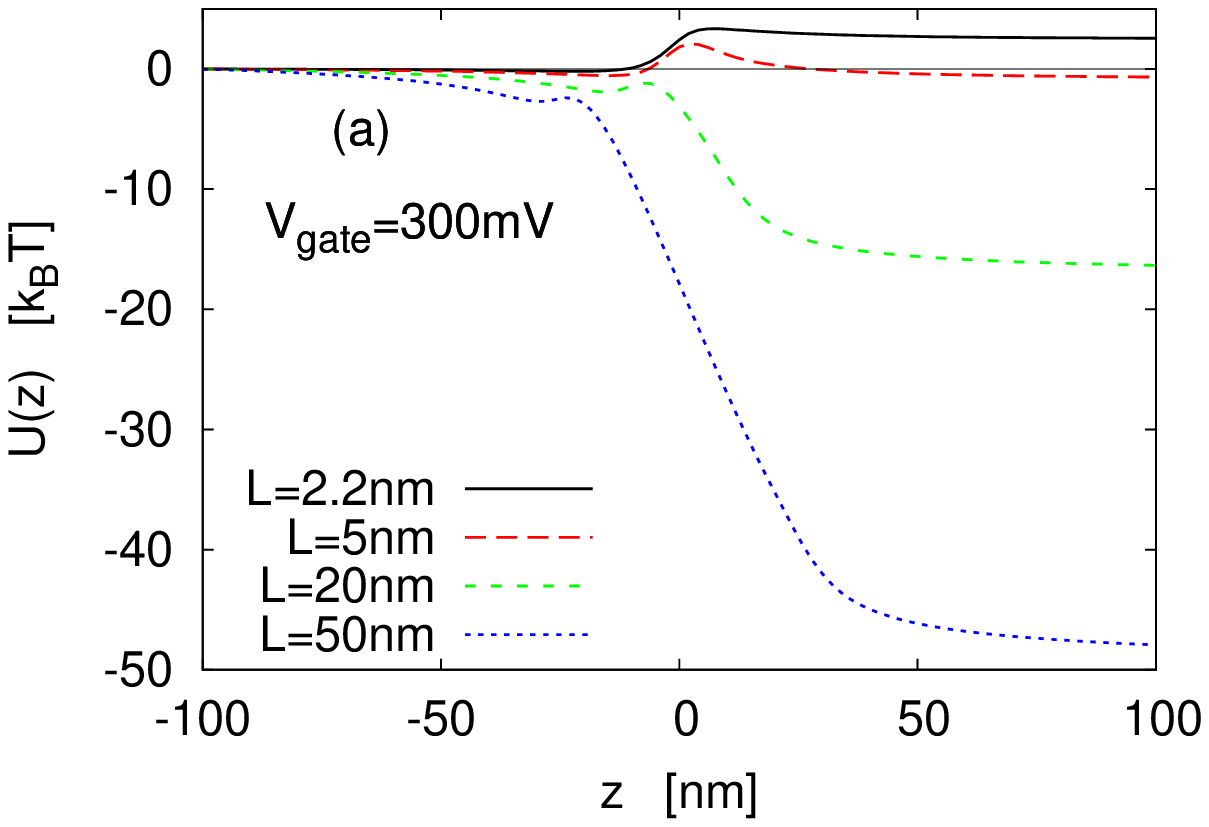}
\epsfxsize=1.0\columnwidth
	\epsfbox{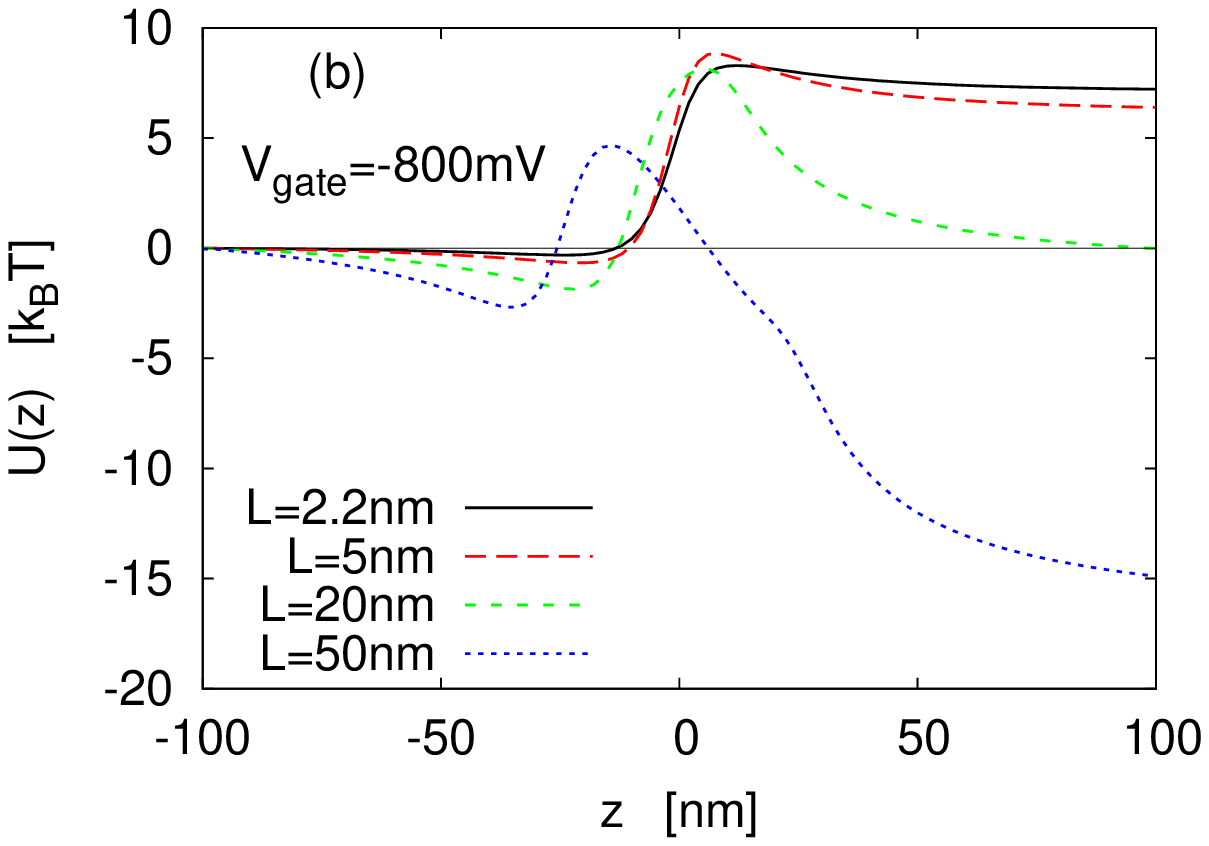}
%	\epsfxsize=0.8\columnwidth
%	\epsfbox{figure4b.eps}
\end{center}
\caption{\label{fig4}
(Color online) Same as in Fig. \ref{fig3a}
($V=100\,$mV, $c_0=100\,$mM,
$\sigma_m=-40\,$mC/m$^2$)
%but for an externally applied voltage $V=100\,$mV
%and an increased membrane surface charge 
%density of $\sigma_m=-40\,$mC/m$^2$.
%Moreover, there is a 
but now for an additional
membrane gate electrode 
of thickness $G=16\,$nm (see Fig. \ref{fig1})
with gate voltage $V_{gate}=300\,$mV (a)
and $V_{gate}=-800\,$mV (b).
}
\end{figure}

The most important consequence of this
observation is illustrated by Fig. 
\ref{fig4}, namely that
the threshold-length, above and below
which particles preferentially move into 
opposite directions, is at about $L=5\,$nm
for $V_{gate}=300\,$mV and can
be moved to about $L=20\,$nm by decreasing the
gate voltage to $V_{gate}=-800\,$mV.
As expected, one finds that also a large variety 
of other threshold-lengths can be realized
by suitably choosing $V_{gate}$, and that analogous 
results are recovered also for other values of 
the surface charges $\sigma_p$ and $\sigma_m$
than in Fig. \ref{fig4}.

\section{Summary and Conclusions}
\label{s6}
%
%For the specific system from Fig. \ref{fig4}
%but without the pore,
%%and for $V=0$, 
%a straightforward caluclation shows that 
%a surface charge density of $\sigma_m=-40\,$mC/m$^2$ 
%and a gate voltage of $V_{gate}=-45\,$mV$+\Delta V$
%generate the same field and charge distributions
%in the ambient fluid as a membrane 
%without gate electrode and a surface charge density 
%of $-40\,$mC/m$^2+\alpha\,\Delta V$ with
%$\alpha=27\,$mC/m$^2$V.
%$-57\,$mV generates exactly 
%the same field and charge distributions
%in the ambient fluid as a membrane 
%without gate electrode, provided 
%%all other system parameters are as in Fig. \ref{fig4}.
%%$\sigma_m=-50\,$mC/m$^2$ in both cases.
%%and the gate electrode is coated with a 
%%$2\,$nm thick non-conducting layer (as
%%in Fig. \ref{fig4}).
%Likewise, a poreless membrane with surface 
%charge density $\sigma_m=-50\,$mC/m$^2+\Delta \sigma$ 
%at $V_{gate}=-57\,$mV$+\alpha\Delta\sigma$
%with $\alpha=46\,$mV\, m$^2$/mC
%is equivalent to an electrodeless 
%membrane with a surface charge density 
%of $-50\,$mC/m$^2$.

%Put differently, Fig. \ref{fig4} illustrates
%how the specific particle length, above and below 
%which the preferential pore passage occurs into 
%opposite directions, can be controlled by 
%means of the gate voltage even when the
%surface charge densities of particle 
%and membrane are quite different.
%In other words, opposite translocations
%of short and long particles is modified
%when the difference between $\sigma_p$ 
%and $\sigma_m$ remains small and disappears
%altogether for larger differences.

In conclusion, we have shown that the 
interplay of hydrodynamic and electrostatic forces 
beyond the realm of small Debye lengths gives
rise to quite non-trivial potential energy 
landscapes, which an elongated cylindrical 
particle encounters while translocating under 
the action of an externally applied voltage
through a solid state nanopore.
In particular, the net potential energy difference 
across the membrane may be of opposite sign for short 
and long particles of constant diameter and charge 
density, for instance DNA-fragments (oligomers) or nano-rods.
Thermal noise thus leads to biased diffusion 
through the pore into opposite directions.
By means of an additional membrane gate electrode
it is even possible to control the specific 
particle length, above and below which the 
preferential pore passage occurs into opposite 
directions.

This new particle sorting
concept seems to us quite worthwhile 
and feasible by today's solid state
nanopore techniques \cite{reviews,gate}.
While the considered setup (Fig. \ref{fig1})
and its quantitative treatment (Poisson, Nernst-Planck, 
and Stokes equations) admittedly neglect many 
details of real systems, we believe 
that our approach still captures 
the essential features quite faithfully.

In other words, the predicted
opposite translocation of long and 
short DNA fragments or nano-rods 
through a solid state nanopore, and its 
control by a gate electrode
seem generic within a wide range of
typical experimental conditions.
%\cite{reviews,he11,nam09,ai10,gra06,lua10}.
Applications e.g. for particle sorting purposes
are immediate, and by employing
many pores in parallel or in series, 
throughput as well as selectivity 
can be readily enhanced \cite{select1,select2}.
%Also with respect to polymer translocation
%through biological protein pores, we expect 
%that analogous, quite intriguing voltage 
%dependent potential energy landscapes will 
%play a key role in future theoretical 
%developments \cite{reviews}.
%More generally, the reported, quite
%intriguing voltage dependent potential
%energy landscapes are also expected
%to substantially advance our theoretical
%understanding of polymer translocation 
%through biological membrane channels 
%\cite{mel03}.

\begin{center}
\vspace{-5mm}
---------------------------
\vspace{-4mm}
\end{center}
This work was supported by Deutsche Forschungsgemeinschaft
under SFB 613 and RE1344/8-1.


\begin{thebibliography}{99}

\bibitem{reviews}
A. Meller,
%{\em Dynamics of polynucleotide transport 
%through nanometre-scale pores},
J. Phys.: Condens. Matter {\bf 15}, R581 (2003);
%Meller, A. 2003.
%Dynamics of polynucleotide transport through nanometre-scale pores.
%J. Phys.: Condens. Matter 15:R581-R607.
%\bibitem{rou05}
%B. Roux, Annu. Rev. Biomol. Struct. {\bf 34}, 153 (2005)
%\bibitem{rou04}
%B. Roux, T. Allen, S. Berneche, and W. Im,
%Theoretical and compuational models of biological ion channels,
%Quaterly Reviews of Biophysics {\bf 37}, 15-103 (2004)
%\bibitem{dek07}
C. Dekker, 
%{\em Solid-state nanopores},
Nat. Nanotech. {\bf 2}, 209 (2007);
%Dekker, C. 2007. Solid-state nanopores. Nat. Nanotechnol. 2:209-215.
D. Branton et al., Nat. Nanotech. {\bf 26}, 1146
%-1153 
(2008);
M. Zwolak and M. Di Ventra, Rev. Mod. Phys. {\bf 80}, 141 (2008);
%S. Howorka and Z. Siwy,
%Nanopore analytics: sensing of single molecules,
%Chem. Soc. Rev. {\bf 38}, 2360
%-2384 
%(2009);
%A. Aksimentiev, Nanoscale {\bf 2}, 468 (2010);
B. M. Venkatesan and R. Bashir, Nat. Nanotech. {\bf 6}, 615 (2011)

%\bibitem{bra08}
%D. Branton et al.,
%The potential and challenges of nanopore sequencing,
%Nat. Biotechnol. {\bf 26}, 1146-1153 (2008)

%\bibitem{ven11}
%B. M. Venkatesan and R. Bashir, Nature Nanotech. {\bf 6}, 615 (2011).
%Venkatesan, B. M., and R. Bashir. 2011. 
%Nanopore Sensors for Nucleic Acid Analysis.
%Nature Nanotech. 6:615-624.

%\bibitem{how09}
%S. Howorka and Z. Siwy,
%Nanopore analytics: sensing of single molecules,
%Chem. Soc. Rev. {\bf 38}, 2360-2384 (2009)

\bibitem{select1}
R. S. Shaw, N. Packard, M. Schr\"oter, and H. L. Swinney, PNAS {\bf 104}, 9580 (2007);
S. M. Iqbal, D. Akin, and R. Bashir, Nat. Nanotech. {\bf 2}, 243 (2007);
D. Goulding, J.-P. Hansen, and S. Melchionna, Phys. Rev. Lett. {\bf 85}, 1132 (2000)


\bibitem{select2}
C.~Marquet, A.~Buguin, L.~Talini, and P.~Silberzan,
%\emph{Rectified motion of colloids in asymmetrically structured channels},
Phys.~Rev.~Lett. {\bf 88}, 168301 (2002);
%
%L.~R.~Huang \emph{et~al.},
%\emph{Role of molecular size in ratchet fractionation},
%Phys.~Rev.~Lett. \bf{89}, 178301 (2002)
%
S. Matthias and F. M\"uller, Nature {\bf 424}, 53 (2003);
%
%L.~R.~Huang, E.~C.~Cox, R.~H.~Austin, and J.~C.~Sturm,
%\emph{Continuous particle separation through 
%deterministic lateral displacement},
%Science {\bf 304}, 987 (2004);
%
%S.~Choi, J.~K.~Park,
%\emph{Microfluidic system for dielectrophoretic 
%separation based on a trapezoidal electrode},
%Lab~on~a~Chip, 2005, \textbf{5}, 1161--1167.
%
%K.~Loutherback, J.~Puchalla, R.~H.~Austin, and J.~C.~Sturm,
%\emph{Deterministic microfluidic ratchet},
%Phys.~Rev.~Lett. {\bf 102}, 045301 (2009);
%
%K.~Loutherback \emph{et~al.},
%\emph{Improved performance of deterministic 
%lateral displacement arrays with triangular posts},
%Microfluid.~Nanofluid., 2010, \textbf{9}, 1143--1149.
%
%P.~Tierno, S.~V.~Reddy, M.~G.~Roper, T.~H.~Johansen, T.~M.~Fischer,
%\emph{Transport and separation of biomolecular cargo on paramagnetic colloidal particles in a magnetic ratchet},
%J.~Phys.~Chem.~B, 2008, \textbf{112}, 3833--3837.
%
G.~Mahmud \emph{et~al.},
%\emph{Directing cell motions on micropatterned ratchets},
Nat. Phys. {\bf 5}, 606 (2009)
%
%L. Bogunovich et al., Soft Matter {\bf 8}, 3900 (2012)

\bibitem{gro10}
A. Y. Grosberg and Y. Rabin, J. Chem. Phys. {\bf 133},
165102 (2010);
M. Wanunu, W. Morrison, Y. Rabin, A. Y. Grosberg, and A. Meller, 
Nat. Nanotechnol. {\bf 5}, 160 (2010)

%\bibitem{elhyd}
%V. A. Parsegian, Nature {\bf 221}, 844 (1969);
%Annu. Rev. Biophys. Bioeng. {\bf 2}, 221 (1973);
%J.~E. Sader and D.~Y.~C. Chan, Langmuir {\bf 16}, 324 (2000);
\bibitem{zha07}
J. Zhang and B. I. Shklovskii, Phys. Rev. E {\bf 75}, 021906 (2007);
%L. Chen and A. T. Conlisk, Biomed. Microdevices {\bf 10}, 289 (2008);

\bibitem{lua08}
B. Luan and A. Aksimentiev, Phys. Rev. E {\bf 78}, 021912 (2008)

\bibitem{rab05}
Y. Rabin and M. Tanaka, Phys. Rev. Lett. {\bf 94}, 148103 (2005)

\bibitem{gho07}
S. Ghosal, Phys. Rev. Lett. {\bf 98}, 238104 (2007)
%\bibitem{gho07} S. Ghosal, Phys. Rev. E {\bf 76}, 061916 (2007).

\bibitem{he11}
Y. He, M. Tsutsui, C. Fan, M. Taniguchi, and T. Kawai,
%Controlling DNA transloxcation through gate modulation of nanopore wall surface charges,
ACS Nano {\bf 5}, 5509
%-5516 
(2011)

\bibitem{sme06}
R. M. M. Smeets, U. F. Keyser, D. Krapf, M.-Y. Wu, N. H. Dekker, and C. Dekker,
Nano Lett. {\bf 6}, 89 (2006); 
H. Chang, F. Kosari, G. Andreadakis, M. A. Alam, G. Vasmatzis, and R. Bashir,
Nano. Lett. {\bf 4}, 1551 (2004)

\bibitem{dor09}
S. van Dorp, U. F. Keyser, N. H. Dekker, C. Dekker, and S. G. Lemay,
Nat. Physics {\bf 5}, 347 (2009)
%S. Das, P. Dubsky, A. van den Berg, and J. C. T. Eijkel, 
%Phys. Rev. Lett. {\bf 108}, 138101 (2012)

\bibitem{bru08}
H. Bruus, {\em Theoretical Microfluidics}, Oxford Univ. Press (2008)

\bibitem{kes11}
S. Kesselheim, M. Sega, and C. Holm, Comput. Phys. Commun. 
{\bf 182}, 33 (2011);
Soft Matter {\bf 8}, 9480 (2012)

\bibitem{fx}
Small (large) Debye lengths 
%$\lambda_D$
can be realized by large (small) ion concentrations \cite{bru08}.
%$c_0$ according to $\lambda_D\propto c_0^{-1/2}$ \cite{bru08}.

\bibitem{similitude}
C. E. Cummings, S. K. Griffiths, R. H. Nilson, and P. H. Paul, 
Anal. Chem. {\bf 72}, 2526 (2000);
J. G. Santiago, {\em ibid.} {\bf 73}, 2353 (2001)
and further references therein.
%M. Firnkes, D. Pedone, J. Knezevic, M. Doblinger, and U. Rant, 
%Nano Lett. {\bf 10}, 2162 (2010);
%J. T. G. Overbeek, in {\em Colloid Science},
%H. R. Kruyt (Ed.), Elsevier, NY 1952;
%F.~A.~Morrison,
%\emph{Electrophoresis of a particle of arbitrary shape},
%J.~Colloid~Interface~Sci. {\bf 34}, 210 (1970);
%E. Yarvi, H. Brenner, and S. Kim,
%SIAM J. Appl. Math. {\bf 64}, 1099 (2004)

\bibitem{yar12}
O. Schnitzer and E. Yariv, Phys. Rev. E {\bf 86}, 021503 (2012)
%E. Yariv, Chem. Eng. Comm. {\bf 197}, 3 (2010)

\bibitem{f1}
As in Ref. \cite{kow11}, we approximate
the pore by a hyperboloid with
$z$-dependent pore radius 
$r(z)=[(\dmin/2)^2+z^2(\dmax^2-\dmin^2)/H^2]^{1/2}$ for
$|z|\leq H/2$. 
In addition, the ``corners'' at $z=\pm H/2$ are
rounded (curvature radius $5\,$nm).
%Also our
%choice $H=20$, $d=20\,$nm, and $D=40\,$nm
%draws on \cite{kow11}
%and references therein.

\bibitem{kow11}
S.~W. Kowalczyk, A.~Y. Grosberg, Y. Rabin, and C. Dekker,
Nanotechnology {\bf 22}, 315101 (2011)

\bibitem{beh01}
S. H.  Behrens and D. G. Grier, J. Chem. Phys. {\bf 115}, 6716 (2001);
D. Stein, M. Kruithof, and C. Dekker, Phys. Rev. Lett. {\bf 93}, 035901 (2004);
M. B. Andersen, J. Frey, S. Pennathur, and H. Bruus, 
J. Colloid Interface Sci. {\bf 353}, 301 (2011)
%F.~H.~J. van der Heyden, D. Stein, and C. Dekker,
%Phys. Rev. Lett. {\bf 95}, 116104 (2005)

%\bibitem{kas85}
%G. Kasper, T. Niida, and M. Yang, 
%J. Aerosol. Sci. {\bf 16}, 535 (1985) 
%and references therein.

\bibitem{she82}
J. D. Sherwood, J. Chem. Soc., Faraday Trans. 2, {\bf 78}, 1091 (1982)

\bibitem{pro03} 
R.~F. Probstein, {\em Physicochemical Hydrodynamics},
Wiley-Interscience, Hoboken, NJ (2003)

\bibitem{ai10}
Y. Ai, J. Liu, B. Zhang, and S. Qian,
%Field effect regulation of DNA translocation through a nanopore,
Anal. Chem. {\bf 82}, 8217
%-8225 
(2010)

\bibitem{che10}
L. Chen and A. T. Conlisk,
Biomed. Microdevices {\bf 12}, 235 (2010)

\bibitem{par73}
V. A. Parsegian, Annu. Rev. Biophys. Bioeng. {\bf 2}, 221 (1973)

%\bibitem{zha12}
%M. Zhang, L. Yeh, S. Qian, J. Hsu, and S.~W. Joo,
%J. Phys. Chem. C {\bf 116}, 4793 (2012)

\bibitem{kow12}
%C.T.A. Wong and M. Muthukumar, J. Chem. Phys. {\bf 133}, 045101 (2010);
S. W. Kowalczyk, D. B. Wells, A. Aksimentiev, and C. Dekker,
Nano Lett. {\bf 12}, 1038 (2012)

\bibitem{jac99}
J.~D. Jackson, {\em Classical Electrodynamics},
Wiley, New York (1999)

\bibitem{par69}
A. Parsegian, Nature {\bf 221}, 844 (1969);
D. J. Bonthuis, J. Zhang, B. Hornblower, J. Math\'e, 
B. I. Shklovskii, and A. Meller, Phys. Rev. Lett. {\bf 97}, 
128104 (2006)

\bibitem{f4}
Regarding Fig. \ref{fig4},
i.e. in the presence of a membrane gate electrode, 
one finds that $\Delta\sigma_m\propto\epsilon_m$ 
(see second to last paragraph).
%and \cite{f5}). 
Apart from this rather trivial 
$\epsilon_m$-dependence, all other effects of 
$\epsilon_m$ and $\epsilon_p$ are again very small.

\bibitem{eis96}  
R.~S. Eisenberg, J. Membr. Biol. {\bf 150}, 1 (1996);
B. Corry, S. Kuyucak, and S. Chung, Biophys. J. 
{\bf 78}, 2364 (2000)
  
\bibitem{mas06}
J.~H. Masliyah and S. Bhattacharjee, 
{\em Electrokinetic and Colloidal Transport Phenomena},
Wiley, NJ (2006);
R. L. Panton, {\em Incompressible Flow},
Wiley, NJ (2005)

\bibitem{f2}
The numerical solution exploits finite element
methods on a triangular grid, properly refined 
within the Debye layers, and typically involving 
about 200'000 
%triangular 
elements. 
%and 5'000 edge elements. 
The basis functions are Lagrange shape
functions of order one 
%(linear) 
for the pressure and
concentration fields, and of order two 
%(quadratic)
for the velocity and electric potential fields.

\bibitem{f3}
We found that the potential energy $U(z)$ 
changes approximately linearly with voltage $V$ 
up to a few $100\,$mV.
%i.e. Figs. \ref{fig2} and \ref{fig3}
%readily yield $U(z)$ for any such $V$.

\bibitem{lub99}
D. K. Lubensky and D. R. Nelson, Biophys. J. {\bf 77}, 1824 (1999)

\bibitem{gate}
P. C. Yen, C. H. Wang, G. J. Hwang, and Y. C. Chou,
Rev. Sci. Instrum. {\bf 83}, 034301 (2012);
Z. Jiang and D. Stein, Phys. Rev. E {\bf 83}, 031203 (2011);
S.-W. Nam, M. J. Rooks, K.-B Kim, and S. M. Rossnagel,
%Ionic field effect transistors with sum-10 nm multiple nanopores,
Nano Lett. {\bf 9}, 2044
%-2048 
(2009);
M. E. Gracheva, A. Xiong, A. Aksimentiev, K. Schulten, G. Timp, and J.-P. Leburton, 
Nanotechn. {\bf 17}, 622 (2006);
B. Luan, H. Peng, S. Polonsky, S. Rossnagel, G. Stolovitzky, and G. Martyna,
Phys. Rev. Lett. {\bf 104}, 238103 (2010);
S. Harrer et al.,  Nanotechn. {\bf 22}, 275304 (2011)



%\bibitem{f5}
%In general, one finds that 
%$\Delta\sigma_m=\epsilon_0\epsilon_m(V_{gate}-\zeta)/h$,
%where $\zeta$ is the 
%%($V$-dependent) 
%zeta-potential \cite{bru08}.
%For our standard parameter values, the latter can be 
%approximately neglected, yielding 
%$\Delta\sigma_m\,\approx \chi V_{gate}$ with $\chi\approx 20\,$mC/m$^2$V.



%\bibitem{lua11}
%B. Luan and A. Aksimentiev,
%J. Phys.: Condens. Matter {\bf 22}, 454123 (2011)

%\bibitem{ai11}
%Y. Ai and S. Qian, Phys. Chem. Chem. Phys. {\bf 13}, 4060 (2011)

%\bibitem{lub99}
%D. K. Lubensky and D. R. Nelson, Biophys. J. {\bf 77}, 1824 (1999);
%P. Reimann, Phys. Rep. {\bf 361}, 57 (2002);
%M. Muthukumar and C. Y. Kong, PNAS {\bf 103}, 5273 (2006)

\end{thebibliography}
\end{document}